\documentclass[12pt]{article}
\usepackage{setspace}
\usepackage{doublespace}
\newcommand{\be}{\begin{equation}}
\newcommand{\ee}{\end{equation}}
\def\bq{\begin{eqnarray}}
\def\eq{\end{eqnarray}}
\def\n{\nonumber}

\def\dex{\delta x}
\def\de{\delta}

\def\de{\delta}

\def\bc{\begin{center}}
\def\ec{\end{center}}

\begin{document}

\renewcommand{\baselinestretch}{2}\pagenumbering{arabic}  \begin{singlespace}
\title{\bf ACTION BASED APPROACH TO THE DYNAMICS OF EXTENDED BODIES IN GENERAL RELATIVITY \footnote{This essay received an ``honorable mention'' in the 2003
Gravity Research Foundation essay competition.}}

\author{Jeeva Anandan$^{\dagger }$, Naresh Dadhich$^{\ddagger }$, Parampreet Singh$^{ \ddagger }$ \\
\small\it${}^\dagger$  Department of Physics and Astronomy, University of South Carolina, \\\small \it Columbia, SC 29208, USA. \\\small \it${}^\ddagger$  Inter-University Centre for Astronomy and Astrophysics, \\\small \it Post Bag 4, Ganeshkhind, Pune~411~007, India. \\{\small \sf email: jeeva@sc.edu, nkd@iucaa.ernet.in, param@iucaa.ernet.in}}
\date{}
\maketitle
\bigskip
\vskip0.5cm
\end{singlespace}
\begin{abstract}We present, for the first time, an action principle 
that gives the equations of motion of an extended body possessing multipole moments in an external gravitational field, in the weak field limit. 
From the action, the experimentally observable quantum phase shifts in the wavefunction of an extended object due to the coupling of its multipole moments with the gravitational field are obtained. Also, since the theory may be quantized using the action, the present approach is useful in the interface between general relativity and quantum mechanics.\end{abstract}

\vskip1cm

Ever since the time of Newton, the motions of bodies, such as an apple or a planet, in the gravitational field has fascinated physicists. After the discovery of general relativity, Einstein, Infeld, and Hoffman \cite{eih} obtained the geodesic equation of motion for an idealized monopole test particle using the local conservation of energy-momentum.
However, real objects, such as astrophysical bodies, are extended and possess multipole moments, in general. 
If the body's extension in space is non negligible compared to the radius of curvature  of the background field then it cannot be treated as a test particle and the modifications to the geodesic equation are expected \cite{dixon1}. 
These modifications are the interaction terms due to the coupling between multipole moments of the body with the background Riemann curvature and its derivatives. Corrections to the geodesic equation due to the coupling of spin \cite{mat,papa} and quadrupole moment \cite{taub,madore,dixon}  to the gravitational field have been obtained using the conservation law for 
energy-momentum tensor. But the corrections due to the coupling of higher multipole moments to the gravitational field have not been obtained previously, to our knowledge.

After the discovery of quantum mechanics, it became necessary to describe the motion of a body by the evolution of its wave function. And the action principle plays a very fundamental role in quantum mechanics. While in classical physics the action is used as a mathematical tool for obtaining the equations of motion by extremization, in quantum mechanics the action is directly observable as the phase of the wave function.  Therefore, if the action is known, it would not only give, on variation, 
the classical equations of motion for all multipoles, but also predict the quantum phase shifts arising out of the coupling of these multipole moments to the gravitational field. For instance, the phase shift in the wavefunction of a molecule due to  the coupling of its dipole moment with gravity, depending upon the available technology, may be measured.  Also, from the action, the evolution of the wave function may be directly obtained in the Feynman path integral formulation of quantum mechanics. However, a procedure to derive equations of motion of extended bodies through an action principle has not been obtained during the past 65 years in which the equations of motion in a gravitational field have been studied.

In this essay we present an action  principle for extended bodies in a gravitational field, in the  weak field limit.
 Our formalism yields in a simple and elegant way the corrections to the geodesic equation for 
{\it all} multipoles of the extended body.  
The equations of motion for all multipoles simply follow from variation of our action, i.e. the corresponding Euler-Lagrange equation. As a demonstration of our formalism we obtain for the first time the force due to the octopole moment. Moreover, our action gives the quantum phase shifts in interferometry due to the coupling of all multiple moments to the gravitational field. The quantum   phase shift due to the coupling of the monopole moment of neutrons with earth's gravitational field has been observed \cite{cow}. With the ongoing experiments in atomic \cite{atomic}, molecular \cite{mol} and Bose-Einstein condensate interferometry \cite{bec}, it 
may be possible in the near future to test the predictions of the phase shifts induced by the multipolar couplings. In this light, our formalism opens up new
 areas for future theoretical and experimental investigations in the interface of the gravitational and quantum realms. 

We envision an extended body as a thin world tube in spacetime, with its thickness small compared to the scale over which curvature varies. Choose in this tube a reference world-line ($z^\mu$) having 4-velocity  $u^\mu = dz^\mu/ds = (1, 0, 0, 0)$ and define multipole moments with respect to it on a spacelike hypersurface as \cite{dixon}
\be
t^{\kappa_1 ... \kappa_n  \mu \nu} = \int \dex^{\kappa_1} ...{\dex^{\kappa_n}} \, \sqrt{-g} \, T^{\mu \nu} \, d^3 x \label{0}
\ee
where  $\de x^\mu = x^\mu - z^\mu$ and $T^{\mu \nu}$ is the energy-momentum tensor. The above multipole moments are defined in a class of coordinate systems that are related by linear transformations in order that the expression (\ref{0}) is covariant. But once they are defined this way,  $ t^{\kappa_1 ... \kappa_n  \mu \nu}$ can be transformed to any arbitrary coordinate system as a tensor. {\it All the relativistic equations in the present essay are covariant with respect to the above linear transformations}, if not with respect to general coordinate transformations. 

Since what we do must be consistent with the Newtonian theory it is important to establish the relation between covariant moments and their Newtonian analogs, which also will give physical meaning to the above multipoles. For that we expand the gravitational potential $\phi(x)$ in a Taylor series around the central world line and hence the potential energy becomes,
\be
U =  \int \, \rho(x) \, \phi(x) \, d^3 x = m \, \phi(z) +  d^i \, \partial_i \phi |_z + \frac{1}{2} \, I^{ij} \, \partial_i \, \partial_j \, \phi |_z +
\frac{1}{6} \, O^{ijk} \, \partial_i \, \partial_j \, \partial_k \, \phi |_z
+ ... 
 \label{eq:U}\ee
where the mass $m = \int \rho(x) \, d^3 x$, the dipole moment $ d^i = \int \, \rho(x) \, \dex^i \, d^3 x = t^{i 00}, $ the quadrupole moment $ I^{ij} = \int \, \rho(x) \, \dex^i \, \dex^jd^3 x = t^{i j 00}$ and the octopole moment
$O^{ijk} = \int \, \rho(x) \, \dex^i \, \dex^jd^3 x = t^{ijk 00}$, with  $\de x^i = x^i - z^i$. The spin tensor in the Newtonian limit is defined as $S^{ij} = 2 \int \rho \, \de x^{[i} v^{j]}$ where $v^i = d x^i/d t$ which upto octopole term satisfies
\be
\frac{d}{dt} S^{ij} = 2 \, p^{[i}  u^{j]} - 2 \, I^{k [i} \partial^{j]} \partial_k \phi|_z - O^{k r [i} \partial^{j]} \, \partial_k \partial_r \phi|_z
\label{eq:spineq}
\ee 
where $u^i = d z^i/d t$ and the momentum $ p^i = \int \, \rho(x) \, v^i \, d^3 x$. The covariantization of the spin tensor leads to $S^{\mu \nu} =  t^{\mu \nu 0} - t^{\nu \mu 0}$. In the weak field Newtonian limit the dipole  term in eq.(\ref{eq:U}) can be written in covariant form as $t^{\mu \alpha \beta}\,  h_{\alpha \beta, \mu}|_z $. Choosing the reference world line as the center of mass, so that  $d^i = 0$, we  write \cite{dixon2}  $t^{\mu \nu \alpha} = S^{\mu (\nu} u^{\alpha)}$, where $S^{\mu \nu}$ satisfies $S^{\mu \nu} u_\nu = 0$. Similarly, the quadrupole term leads to $h_{\alpha \beta, \mu \nu}|_z \, t^{\mu \nu \alpha \beta} = - 2 \, R_{\alpha \mu \beta \nu}|_z \, I^{\mu \nu} \, u^{\alpha} \, u^{\beta}$ and
the octopole term leads to $h_{\alpha \beta, \mu \nu \sigma}|_z \, t^{\mu \nu \sigma  \alpha \beta} = - 2 \, R_{\alpha \mu 
\beta \nu, \sigma }|_z \, O^{\mu \nu \sigma} \, u^{\alpha} \, u^{\beta}$
in the weak field limit. One can in a similar way obtain the relationships between the higher order multipoles and their Newtonian analogs; however
 here we restrict ourselves up to the octopole moment. 

We  now use these relations to obtain the corrections to the geodesic equation through an action principle.  For simplicity we restrict perturbations of the metric to the first order and later covariantize the equation of motion by changing ordinary derivative to covariant derivative. We further neglect the back reaction of the extended body on the background field and choose the coordinates such that along the reference world-line $g_{\mu\nu} = \eta_{\mu\nu}$. The action for the extended body can be written as
\bq
{\cal S} &=& \n \int \sqrt{-g} \, {\cal 
L}|_{g_{\mu \nu} = \eta_{\mu \nu}} \, d^4 x 
+ \int \de g_{\mu \nu} \, \frac{\de}{\de g_{\mu \nu}} (\sqrt{-g} {\cal L}) \, d^4 x + ... \\ &=&   - \,  \int p_{\alpha} \, u^{\alpha} \, d s + \frac{1}{2} \, \int \de g_{\mu \nu} \, \sqrt{-g} \, T^{\mu \nu} \, d^4 x + ... \label{eq:var}
\eq
where the first term is the kinetic energy term. Since $\de g_{\mu \nu} = g_{\mu \nu} - \eta_{\mu \nu}$, it may be expanded in the Taylor series as $\de g_{\mu \nu} =  h_{\mu \nu, \sigma}|_z \, \de x^{\sigma} + \frac{1}{2} \,h_{\mu\nu, \sigma \rho}|_z \, \de x^{\sigma} \, \de x^{\rho} + \frac{1}{6} \, h_{\mu\nu, \sigma \rho \alpha}|_z \, \de x^{\sigma} \, \de x^{\rho} \, \de x^{\alpha}..., $ where we have  $h_{\mu \nu}(z) = 0$. In the present  weak field limit, we neglect all terms that are quadratic or higher order in $\de g_{\mu \nu}$. Substituting this in  eq.(\ref{eq:var}) and using the relationship between the spin tensor, quadrupole moment, octopole moment and the covariant multipoles, we get
\bq\label{eq:action}
{\cal S}  &=& \n  - \,  \int p_{\alpha} \, u^{\alpha} \, d s + \frac{1}{2} \, \int h_{\alpha \beta, \mu} \, S^{\mu \alpha} \, u^{\beta} \, d s \\
&~~& - \frac{1}{2} \,\int \,  R_{\alpha \mu \beta \nu} \, I^{\alpha \beta} \, u^{\mu} \, u^{\nu} \, d s  
- \frac{1}{6} \, \int \,  R_{\alpha \mu \beta \nu, \rho} \, O^{\alpha \beta \rho} \, u^\mu \, u^\nu \, ds +....,
\eq 
upto octopole moment.
The equation of motion for an extended body can now be simply obtained by extremizing the action subject to coordinate variations vanishing at the end points of the path. 
The equation of motion containing the forces on the body till the octopole moment can hence be obtained to be, 
\be
\frac{D p^*_{\sigma}}{Ds} =  \frac{1}{2} \, R_{\sigma \lambda \mu \nu} \, 
u^{\lambda} \, S^{\mu \nu} + \frac{1}{6} \, J^{\mu \alpha \beta \nu} \, 
\nabla_{\sigma} R_{\mu \alpha \beta \nu} + \frac{1}{6} \nabla_\sigma \nabla_\rho \, R_{\alpha \mu \beta \nu}\,  u^{[\alpha} O^{\mu]\rho [\nu} u^{\beta]}
\ee 
where in order to compare with earlier results obtained till
quadrupole \cite{dixon}, we have defined 
 $p^*_{\sigma} = p_\sigma - R_{\sigma \lambda \mu \nu} \, I^{\lambda \mu} \, u^\nu - \frac{1}{3} \nabla_\rho R_{\sigma \mu \nu \alpha} O^{\mu \nu \rho} u^\alpha  $ and $J^{\mu \alpha \beta \nu} := - 3 \, u^{[\mu} I^{\alpha ][\beta} u^{\nu ]}$. 
Similarly, the spin propagation equation, eq.(\ref{eq:spineq}), yields,
\be\label{7}
\frac{D}{Ds} S^{\alpha \beta} = 2 \, p^{* [\alpha} \, u^{\beta]} - \frac{4}{3} \, R^{[\alpha }{}_{\mu \nu \sigma} \, J^{\beta] \mu \nu \sigma} - R^{[\beta}{}_{\mu \nu \sigma;\rho} \, O^{\alpha] \rho [\sigma } \, u^{\nu]} u^\mu ~~.
\ee
Thus we have derived for the first time  propagation equations for an extended 
body with the octopole moment. Previous results which were obtained till
quadrupole are consistent with the above equations.
 Here we have established a very simple and elegant method based on action principle which can be used to derive higher order corrections due to any multipole moment.

This approach also predicts the resulting quantum phase shifts due to the coupling of multipole moments of an extended body with the background gravitational field.  In the WKB approximation, the phase $\phi={{\cal S}\over\hbar}$, where ${\cal S}$ is given by (\ref{eq:action}), i.e.
\bq\label{eq:phase}\phi  &=& \n  - \,  {1\over\hbar}\int p_{\alpha} \, u^{\alpha} \, d s + \frac{1}{2\hbar} \, \int h_{\alpha \beta, \mu} \, S^{\mu \alpha} \, u^{\beta} \, d s \\
&~~& - \frac{1}{2\hbar} \,\int \,  R_{\alpha \mu \beta \nu} \, I^{\alpha \beta} \, u^{\mu} \, u^{\nu} \, d s  -\frac{1}{6\hbar} \, \int \,    R_{\alpha \mu \beta \nu, \rho} \, O^{\alpha \beta \rho} \, u^\mu \, u^\nu \, ds +....
\eq
The monopole term in the phase (\ref{eq:phase}) has been experimentally observed before in the non relativistic limit \cite{cow,jeeva1}.  The  second term in (\ref{eq:phase}) is in agreement, in the present linearized limit, with the phase shift due to the coupling of  intrinsic spin of the neutron to the gravitational field obtained previously \cite{jeeva4}.  But the present result extends the latter result to include orbital angular momentum, which makes it much bigger for Bose-Einstein condensates.  The quadrupole, octopole and the higher order terms, which can easily be computed in (\ref{eq:phase}) using the above procedure, are completely new. We hope that interferomoteric experiments with quantum extended objects will verify these new gravitationally induced phase shifts, in the near future.

Acknowledgements: JA thanks an NSF and ONR grants for partial
support and PS thanks 
CSIR for a research grant. 

\begin{singlespace}

\end{singlespace}
\end{document}